\documentclass[aps,preprintnumbers,amsmath,amssymb,floatfix,groupedaddress,nofootinbib]{revtex4}

\usepackage{enumerate}
\usepackage{graphicx}
\usepackage{color}
\usepackage[utf8]{inputenc}
\usepackage{enumitem}
\usepackage{appendix}
\usepackage{ulem}

\usepackage{fontenc}  
\usepackage{amsmath}

\newcommand{\beq}{\begin{equation}}
\newcommand{\eeq}{\end{equation}}
\newcommand{\bea}{\begin{eqnarray}}
\newcommand{\eea}{\end{eqnarray}}
\newcommand{\ba}{\begin{array}}
\newcommand{\ea}{\end{array}}

\newcommand{\bef}{\begin{figure}}
\newcommand{\eef}{\end{figure}}

\begin{document}

\title{Kolmogorovian Censorship, Predictive Incompleteness, \\
and the locality loophole in Bell experiments.}

\author{Philippe Grangier}
\affiliation{Laboratoire Charles Fabry, IOGS, CNRS, 
Universit\'e Paris Saclay, F91127 Palaiseau, France.}

\begin{abstract}

We revisit the status of quantum probabilities in light of Kolmogorovian Censorship (KC) and the Contexts, Systems and Modalities (CSM) framework, and we compare KC‑based frameworks  with alternatives such as superdeterminism, supermeasurements, and predictive incompleteness. After briefly recalling the technical content of KC and its scope, we show that KC correctly identifies that probabilities are classical within a fixed measurement context but does not by itself remove the conceptual tension that motivates nonlocal or conspiratorial explanations of Bell‑inequality violations. We argue that predictive incompleteness — the view that the quantum state is operationally incomplete until the measurement context is specified — provides a simple, minimal, and explanatory framework that preserves relativistic locality while matching experimental practice. Finally we clarify logical relations among these positions, highlight the assumptions behind them, and justify the move from Kolmogorov's to Gleason's framework for quantum probabilities.

\end{abstract}

\maketitle

\section{Introduction}

In the scientific literature there are still debates to decide whether quantum mechanics (QM)  is contextual \cite{ContextualReview,ContextualPG} or noncontextual \cite{histories}, and whether quantum probabilities fit within the usual Kolmogorovian  framework \cite{LZ1,LZ2} or not  \cite{holik,svozil,book,andrei}. For the sake of completeness, Kolmogorov's axioms are briefly presented in Appendix A. 
To some extent, these debates can be considered as matters of definitions, so it is useful  to spell out the explicit or implicit assumptions that are behind these different views. 
A milestone in this debate is the Kolmogorovian Censorship (KC) \cite{LZ1,proof,redei}, stating  that quantum probabilities can be identified with classical, Kolmogorovian probabilities when the measurement context has been specified, 
or more generally when a classical probability distribution of measurement contexts has been given. 
The KC sounds rather obvious on a physical intuitive basis, and technically  it has been demonstrated for a countable number of measurement contexts \cite{proof,redei}. It also has some relationship with superdeterminism or supermeasurements of Refs. \cite{Sabine,Palmer} as it will appear more clearly below. 

\section{Kolmogorovian Censorship and Bell experiments.}

Looking in more details, there is some controversy on the physical meaning of the KC, as we will explain now.  In particular, Szab\'o et al \cite{LZ1,LZ2}  claim that  QM probabilities are Kolmogorovian, not only in a single context, but also in a loophole-free Bell experiment, where a random choice among four different contexts is implemented \cite{aa}. Then they conclude that Bell's inequalities are either irrelevant (when considering the four contexts separately), or not violated, when considering the four contexts together: since each one has a 1/4 probability to occur,   Bell's $S$ parameter (see Appendix B)
is reduced from $2 \sqrt{2}$ down to  $\sqrt{2}/2$, that is below 2 as it would be expected classically. 

More precisely, it is uncontroversial that the probabilities predicted by QM in a given context, or commuting subalgebra, are Kolmogorovian; the problem arises when trying to ``gather" probabilities predicted by QM in different contexts. Said otherwise, in any given context it is possible to build a Kolmogorovian probability distribution, or equivalently a hidden variable theory; however this distribution is matched to 
the actual context. 

In a loophole-free Bell test \cite{aa}, the remote random choices of measurements (polarizers orientations) are designed to forbid that this matching may be done by a relativistically causal transmission between the source and the detectors. Despite these barriers, the matching might be obtained either by an instantaneous influence between the source and the detectors (explicit nonlocality), or by assuming that it is pre-established before the actual experiment takes place (super-determinism). 
One has then the following options:

\begin{enumerate}

\item Szab\'o et al \cite{LZ1,LZ2}  claim that the only relevant way to interpret QM probabilities in different contexts is to consider a classical probability distribution over these different contexts. This corresponds by construction to a feasible experiment, typically a loophole-free Bell experiment, including the random choice among four different contexts \cite{aa}.  Then the global probability  distribution is Kolmogorovian, and Bell's inequalities (BI) are not violated, because as written above  the probabilities in each context are divided by four, as well as the resulting $S$ value. However a drawback of this approach is that the pre-established matching between the source and the measurements is still required, through some kind of ``global determinism" \cite{durt}. Therefore, though BI are not explicitly violated, the basic physical problem of the origin of the matching is still present. 

\item 
Rather than considering a probability distribution over the contexts, one may calculate  the correlation function in each context, which is the very idea of a loophole-free Bell test, and gather them in Bell's $S$ value. However, one may argue that the four correlations functions correspond to four different incompatible experiments, and thus bringing their results together is counterfactual, because they  cannot be measured simultaneously; then BI cannot be demonstrated. This is a standard answer to the problem, 
but it remains debated; see below for further explanations. 

\item Still an alternative way is to admit that the results in the four contexts can be combined, as it would be the case classically since they apply to the same system; then one gets Bell's inequalities, which are experimentally violated, so one should explain why. We consider that this is a meaningful question, and there are basically three options, spelled out in detail in \cite{entropy} (see also Appendix C):

\hspace{-6mm} 3a, 3b - in the spirit of option 1 above,  
the way to violate Bell's inequalities is by admitting that the system's parameters and the orientations of the polarizers are not independent variables, despite the fact that these orientations are chosen randomly and independently at a large distance. This can be obtained either by (3a)~admitting superdeterminism i.e. denying the possibility of independent random choices \cite{durt}, or by (3b)~
admitting a non-local influence between the source and the measurements \cite{entropy}. These two options have recently been shown to be equivalent \cite{blasiak}, and they are in our opinion equally undesirable - though they are matter of ontological choice, and cannot be proven wrong.  

\hspace{-6mm} 3c - the third option, known as predictive incompleteness \cite{entropy,jarrett}, is to recognize that the quantum state by itself is not enough to specify  the measured probability distribution, as long as the context has not be specified. This contradicts predictive completeness (also called outcome independence), which is a required hypothesis for Bell's theorem, and therefore  BI cannot be demonstrated.  This conclusion corresponds to the detailed analysis presented in \cite{entropy}, and also the more general framework presented in \cite{completing} to ``complete" the usual quantum state by specifying the measurement context. These two papers, as well as the arguments above, are consistent within  the general quantum framework called CSM (Contexts, Systems and Modalities) \cite{CSM1,CSM2}. 
\end{enumerate}

\section{Discussion: from Kolmogorov's axioms to Gleason's theorem.}

\subsection{Kolmogorovian censorship vs predictive incompleteness.}
From the above it should be clear that the ontologies underlying either Szab\'o's position or CSM are quite different.  Szab\'o et al claim that QM probabilities are Kolmogorovian and that QM can ultimately be seen as a (super)deterministic theory. On the other hand, CSM  stipulates that QM is fundamentally non-deterministic, due to the conjunction of quantization and contextuality \cite{Gleason,Uhlhorn}, and that quantum probabilities are non-Kolmogorovian, unless restricted to a single context according to the KC. 
 \vskip 2mm
 
This is why the (non-Kolmogorovian) predictive incompleteness of $\psi$ is useful: it leaves enough freedom so that the independent choice of the measurement can contribute to the determination of the observed probability distributions, avoiding both nonlocality and superdeterminism. It  is worth emphasizing again, as spelled out in \cite{entropy}, that predictive incompleteness makes no sense in classical physics, and appears as a specific quantum feature - thereby able to consistently explain the violation of Bell's inequalities. 

\subsection{What about non-measurable subsets ?}
Let’s note that the main proof in \cite{LZ2} is based on a theorem in Pitowsky’s book  \cite{Pbook}, which speaks also about the violation of Bell’s inequalities using non-measurable  subsets of local hidden variables (Chapter 5 in \cite{Pbook}). This kind of ideas were developed more recently \cite{Sabine,Palmer}, and are very controversial \cite{Scott,Matteus}. We don’t want to enter in this debate here, but we note the following sentence by Palmer \cite{Palmer}: ``Importantly, this means a non-conspiratorial interpretation of (our approach) implies that physical theory does not have the post-hoc property of counterfactual definiteness. (…) This important point seems to have been lost in more recent discussions of Bell’s Theorem.''
 \vskip 2mm
 
This gives a major hint about the meaning of Palmer’s approach:  if  counterfactual definiteness 
- that is, treating in statistical calculations the results of unperformed measurements on an equal footing as actual results  - is rejected, then Bell’s inequalities don’t hold any more (option 2. above); this is well known  and has not ‘been lost in more recent discussions of Bell’s Theorem’. Then the work in \cite{Sabine,Palmer} may be seen as a possible way to justify a rejection of counterfactual definiteness  \cite{Hance}, but there is a much simpler one : counterfactual definiteness is not compatible with predictive incompleteness, see Appendices and  \cite{entropy}.  Actually Ref. \cite{Sabine} gives a good definition of predictive incompleteness: it is ``the lack of information about the measurement outcomes in the wave-function, combined with the fact that an observation in one location can tell us something about the measurement outcome in another location". Then \cite{Sabine} is trying to explain this by using supermeasurements, but  here we get it  without using any non-measurable set of hidden variables - as a consequence of the CSM contextual quantization postulates \cite{CSM1,CSM2,Gleason,Uhlhorn}. 
Again, acknowledging that the quantum state vector is  predictively incomplete may appear disturbing, but it makes perfect sense and ensures that many ``paradoxes" are removed in the CSM framework.

\subsection{From $\sigma$‑algebras to projection lattices: Gleason's route and the gluing of contexts}

A compact way to understand the formal passage from classical to quantum probability is to observe that the classical domain of events --- a Boolean $\sigma$‑algebra, see Appendix  --- is replaced in quantum theory by the lattice $\mathcal P(\mathcal H)$ of orthogonal projections on a Hilbert space $\mathcal H$. In the classical Kolmogorov framework additivity is required for disjoint sets; Gleason's theorem implements the exact analogue of this requirement in the quantum setting by demanding additivity for \emph{mutually orthogonal} projections. 

Concretely, one considers a function
\(
m:\mathcal P(\mathcal H)\to[0,1]
\)
satisfying
\begin{enumerate}
  \item $m(I)=1$ (normalization),
  \item if $P_iP_j=0$ for $i\neq j$ then \(m\!\bigl(\sum_i P_i\bigr)=\sum_i m(P_i)\) (orthogonal additivity).
\end{enumerate}
Gleason's theorem shows that, for $\dim\mathcal H\ge3$, any such $m$ is of the trace form,

\[
m(P_i)=\operatorname{Tr}(\rho P_i),
\]
for a unique density operator $\rho$. Thus the formal move of replacing a classical $\sigma$‑algebra by the projection lattice and replacing disjoint additivity by orthogonal additivity — is precisely the route Gleason follows. This perspective clarifies the sense in which contexts are ``glued.'' A single measurement context corresponds to a maximal commuting family of projections; restricted to that commuting subalgebra the projection lattice is isomorphic to a classical Boolean algebra and the probabilities are Kolmogorovian. 

The nontrivial content of Gleason's hypotheses  is to ask for a single assignment $m$ that is consistent across all such commuting subalgebras simultaneously: 
the value assigned to a projection does not depend on which orthogonal decomposition (which context) it appears in. The essential fact that a given projector appears in a continuous infinity of context, provided that $\dim\mathcal H\ge3$, is called intertwining of contexts by Gleason. In the CSM framework, a modality is defined as the association of a projector $P_i$ (a usual quantum state) and a context (a complete set of orthogonal projectors including $P_i$). Sharing a projector is then an equivalence relation for modalities; it is called extravalence \cite{CSM2}, and corresponds physically to mutual certainty. This justifies that the probability depends only on the projector (the extravalence class) and not on the embedding context. 

Two additional remarks are in order. First, the lattice $\mathcal P(\mathcal H)$ is non‑Boolean (non‑distributive), and this structural difference is the source of contextuality: additivity on orthogonal families is a strictly different requirement from full $\sigma$‑additivity on a Boolean algebra. Second, Gleason's theorem requires $\dim\mathcal H\ge3$; nevertheless, the two‑dimensional case is included, 
provided that it is embedded into higher dimensions \cite{Gleason}. It is actually well known that one can build a classical model for a single qubit, but not for two qubits. 

Operationally, Kolmogorovian Censorship and Gleason's hypotheses are therefore complementary: KC explains why probabilities look classical inside a fixed context, while Gleason characterizes when and how those context‑by‑context classical assignments can be consistently extended to a single quantum probability law on the whole projection lattice.

\subsection{A philosophical detour through physical realsim.}

We have seen that the KC in a single context is easily integrated in the CSM framework, but one should not conclude that quantum probabilities are Kolmogorovian in a classical sense: this would be true in classical physics, because there is only one universal context, but this fails in quantum physics, because there is a continuous infinity of different, incompatible contexts. This point of view is implicit in textbook quantum mechanics, and it is made explicit in the CSM framework, by using operator algebra and infinite tensor products. This framework allows a contextual unification of classical and quantum physics, within a unique macroscopic physical world  \cite{ContextualPG,MP1,MP3,enigma}.
 \vskip 2mm
 
 One may notice that R\'edei \cite{redei}  (see also \cite{gomori}) considers that the KC is problematic, in particular because  it implies that ``probabilities are thus not features of quantum systems in and of themselves, they are features that only manifest themselves upon measurement. Philosophers (or physicists) with a robust realist conviction may find unattractive this strongly instrumentalist flavor of interpretation of quantum probability forced upon us by the KC.'' In the CSM approach, quantum probabilities also get a meaning only upon measurement, due to the predictive incompleteness of $\psi$, and they are also genuinely non-classical probabilities.  Nevertheless, CSM is 
 based on physical realism, as the statement that {\it the purpose of physics is to study entities of the natural world, existing independently from any particular observer's perception, and obeying universal and intelligible rules} \cite{completing,CSM1,CSM2}.  
  We claim therefore that even with a ``robust realist conviction" one can accept that $\psi$ is predictively incomplete - as the best way to make sense of this conundrum, within the framework of contextual objectivity \cite{CO2002}.

\section{Conclusions.}
\vspace{-2mm}
The Kolmogorovian Censorship is a useful technical observation: quantum probabilities are Kolmogorovian when a measurement context is fixed, but this formal fact does not resolve the deeper interpretational choice exposed by Bell‑type experiments. Superdeterminism and nonlocal hidden‑variable accounts restore a single global joint distribution only at the cost of questionable ontological commitments and scientific methodology. By contrast, predictive incompleteness accepts that the quantum state alone does not determine outcome statistics across incompatible contexts and thereby preserves locality and empirical adequacy without invoking conspiratorial correlations. It also leads straightforwardly to Gleason's theorem and Born's rule.  Framing the debate in these terms clarifies which extra assumptions are being made in different reconstructions and points to concrete empirical and conceptual criteria that future work should address.
\vskip 2mm

As a final remark, a significant challenge in quantum  reconstructions is to state clearly what are the hypotheses or postulates, and what are their consequences. It should also been made clear whether one looks for a fully deductive reasoning, or a partially inductive one, that is  an ``Inference to the Best Explanation''  \cite{Gleason,Uhlhorn}. In the CSM point of view a fully deductive approach does not fit, since, quoting Landau  \cite{landau}, ‘quantum mechanics  (...) contains classical mechanics as a limiting case, yet at the same time it requires this limiting case for its own formulation’. Then what is desired is not a deduction of the Laws of Nature from some postulates in a mathematical sense, but rather a fully consistent construction, including both classical and quantum physics from the beginning, and clearly separating 
experimentally based evidence from its mathematical description \cite{ContextualPG,MP1,MP3,enigma}. 
\vskip 2mm

\acknowledgments{P.G. thanks Mathias Van Den Bossche, Olivier Ezratty, Franck Lalo\"e, Roger Balian, and Herv\'e Zwirn for interesting discussions. Copilot has been used to edit this document. }

\appendix

\section{Classical Kolmogorov framework.}
\vspace{-2mm}

This Appendix  provides a short, self‑contained account of the classical Kolmogorov probability framework. We start with the basic mathematical objects, then state the axioms, and finally give some general remarks. 

\subsection{Basic set‑algebraic objects.}

\textbf{(1) $\sigma$‑algebra.} Let $\Omega$ be a nonempty set. A family of subsets $\mathcal F\subset 2^\Omega$ is a \emph{$\sigma$‑algebra} (tribu) if (i) $\Omega\in\mathcal F$, (ii) $A\in\mathcal F\Rightarrow A^c\in\mathcal F$, and (iii) $\{A_n\}_{n\ge1}\subset\mathcal F\Rightarrow\bigcup_{n\ge1}A_n\in\mathcal F$. Closure under countable unions (and hence countable intersections) is the defining extra structure that distinguishes a $\sigma$‑algebra from weaker set algebras.
\vskip 2mm 

\textbf{(2) Borel $\sigma$‑algebra.} If $(X,\tau)$ is a topological space, the \emph{Borel $\sigma$‑algebra} $\mathcal B(X)$ is the smallest $\sigma$‑algebra containing all open sets (equivalently all closed sets). It is the canonical measurable structure tied to topology and is the usual domain for continuous or Borel‑measurable observables. One often passes to the \emph{completion} of $\mathcal B(X)$ with respect to a given measure to include all null subsets.
\vskip 2mm 

\textbf{(3) Boolean algebra of sets.} A family $\mathcal A\subset 2^\Omega$ is a \emph{Boolean algebra} if it is closed under finite unions, finite intersections and complementation (equivalently under symmetric difference and intersection). Every $\sigma$‑algebra is a Boolean algebra, but a Boolean algebra need not be closed under countable operations; conversely, the $\sigma$‑algebra generated by a Boolean algebra is obtained by closing it under countable unions and complements.
\vskip 2mm 

{\bf Why $\sigma$‑algebras for probability.}
Kolmogorov's modern axioms require a domain that supports \emph{countable additivity} and limit operations (monotone convergence, dominated convergence, product measures, etc.). Finite closure (Boolean algebra) is insufficient for these analytic tools. Thus probability theory is formulated on $\sigma$‑algebras.

\subsection{Kolmogorov axioms in a measure‑theoretic form.}

In order to state the Kolmogorov axioms, one needs 
a sample space $\Omega$, which is the set of all possible outcomes or elementary events.
The space $\mathcal F$ of all events, taken to be subsets of  $\Omega$, 
must be a $\sigma$-algebra on $\Omega$, as explained above. 
Then the probability measure $P$ assigns to each event $A \in F$
its probability $P(A)$ with the properties 

\begin{enumerate}
\item \textbf{Positivity:} $P(A) \geq 0$.
  \item \textbf{Normalization:} $P(\Omega)=1$.
  \item \textbf{Countable additivity:} For any countable sequence of pairwise disjoint events $(A_n)_{n\ge1}\subset\mathcal F$,
\[
    P\Bigl(\bigcup_{n\ge1}A_n\Bigr)=\sum_{n\ge1}P(A_n).
  \]
\end{enumerate}
From these axioms follow the usual consequences (finite additivity, continuity from above and below, monotonicity). This is the standard modern formulation used throughout probability theory.

\subsection{Relations and practical remarks.}

\begin{itemize}
  \item \textbf{Generation:} Given any collection $\mathcal C\subset 2^\Omega$ (for instance a Boolean algebra of "elementary" events), there is a smallest $\sigma$‑algebra $\sigma(\mathcal C)$ containing $\mathcal C$; measures are then defined on $\sigma(\mathcal C)$. In practice one often starts from a Boolean algebra of operational events and passes to the generated $\sigma$‑algebra to enable countable additivity.
  \item \textbf{Borel vs arbitrary $\sigma$‑algebra:} The Borel $\sigma$‑algebra is natural when topology matters (observables continuous or Borel measurable). The full power set $2^\Omega$ is usually too large (non‑measurable pathologies); working with Borel sets or its completion avoids these issues.
  \item \textbf{Completion:} Given $(\Omega,\mathcal F,P)$, the \emph{completion} $\overline{\mathcal F}$ adds all subsets of $P$‑null sets; completed spaces are convenient for almost‑sure statements and avoid dependence on particular representatives of null classes.
  \item \textbf{Random variables and measurability:} A map $X:(\Omega,\mathcal F)\to (S,\mathcal S)$ is a random variable if $X^{-1}(B)\in\mathcal F$ for every $B\in\mathcal S$. When $S=\mathbb R$ and $\mathcal S=\mathcal B(\mathbb R)$ this is the usual Borel measurability requirement.
  \item \textbf{Product spaces:} Product $\sigma$‑algebras (generated by measurable rectangles) are essential for constructing joint distributions and for independence; countable closure is again crucial here.
\end{itemize}

\noindent {\bf Algebraic summary.}
A \emph{Boolean algebra} models finite logical combinations of events; a \emph{$\sigma$‑algebra} is the countably closed extension required for measure theory; the \emph{Borel $\sigma$‑algebra} is the canonical choice when topology is present; Kolmogorov probabilities are measures on $\sigma$‑algebras satisfying normalization and countable additivity, and most practical constructions use Borel (or completed Borel) $\sigma$‑algebras as the measurable domain.

In quantum mechanics the axioms of positivity, normalization and countable additivity are still true, but the global event space in no more a $\sigma$-algebra.  This has important consequences discussed in the main text.

\section{Bell test with random switching.}

%
Consider a Bell test with fast random switching of the polarizers orientations \cite{aa} between $x$ and $x'$ on one side, $y$ and $y'$ on the other side, where the results $A(x)$,  $A(x')$, $B(y)$, $B(y')$ are denoted $\pm 1$. The four correlation coefficients like $E(x,y) = \overline{A(x) B(y)}$ are calculated from the number of counts $N_\pm(x,y)$, with similar expressions for $(x',y), (x,y'), (x',y')$, which are all gathered together during the experiment.  Then Bell's inequalities tell that $| S | \leq 2$, where  $S = E(x,y) +  E(x',y) + E(x',y') - E(x,y')$, in conflict with QM that predicts $S_{max} = 2 \sqrt{2}$. 

On the other hand, if the normalization of  the correlation coefficients is done by summing all counts for all possible orientations and results, 
one may consider that $x$ and $y$ are also issued from independent random processes in variable spaces $X$ and $Y$, as it is done in loophole-free Bell tests \cite{aa}. Then the global probability writes
\beq
P(x a , y b) = \sum_{\lambda \in \Lambda, x \in X, y \in Y} P(ab|xy\lambda) P(\lambda) P(x) P(y) \label{eqs1}
\eeq
where $P(ab|xy\lambda)$ is given by standard formulas \cite{entropy}. 
Taking $\Lambda = \{\lambda \}$, $X=\{x, x'\}$ with $P(x)=P(x')=1/2$, 
$Y=\{y, y'\}$ with $P(y)=P(y')=1/2$  as in a usual Bell test, 
one gets  $P(x a , y b) =  P(ab|xy\lambda)/4$.  

Correspondingly, the random variable $(x a , y b)$ may take 16 mutually exclusive values, not 4, and Bell's inequalities cannot be written anymore. 
This is because Bell's reasoning requires to calculate  the correlation functions $E(x,y) = \langle a b \rangle_{x,y}  $ by using $P(ab | xy)$, not $P(x a , y b)$, so that the four different measurements apply to the same sample space $\Lambda$. This means implicitly that  $\lambda$ completely carries the pairs' properties, and the measurement results can be predicted from the knowledge of $\lambda$ alone, as it would be the case in classical physics. But this is counterfactual 
with respect to the quantum approach, where $\{\lambda,x,y\}$, $\{\lambda,x,y'\}$, $\{\lambda,x',y\}$, $\{\lambda,x',y'\}$ are four different situations  
which should not be merged within a single $S$ value, contrary to Bell's reasoning \cite{Bell,BCH}. This is another way to tell that $\psi$ is predictively  incomplete, and requires a context specification to be turned into an actual probability distribution. 

\vspace{-2mm}
\section{Three options for violating Bell's inequalities.}

\begin{itemize}
\item \vspace{-2mm} \textbf{Superdeterminism (option 3a)} 
\begin{itemize}
\item \textbf{Core idea:}   Measurement settings are not independent of hidden variables; pre-existing correlations determine outcomes and choices. 
\item \textbf{Costs / risks:} Undermines statistical independence/free-choice assumption; risks unfalsifiability if correlations are unconstrained; experimental physics becomes conceptually fragile. 
\item \textbf{Benefits / implications:} Restores locality in principle; can reproduce quantum statistics with appropriate hidden-variable dynamics. 
\end{itemize}

\item \textbf{Nonlocality (option 3b)}
\begin{itemize}
\item \textbf{Core idea:}  Outcomes at one wing depend on spacelike-separated settings/outcomes at the other. This dependance may occur only at the level on non-accessible hidden variables (de Broglie-Bohm theory). 
\item \textbf{Costs / risks:} Tension with relativistic causality and signal locality, even if “no-signalling” is preserved; conceptual burden of action-at-a-distance. 
\item \textbf{Benefits / implications:} Explains Bell inequality violations within hidden-variable models; widely used narrative in quantum foundations, often backed up by de Broglie-Bohm theory. 
\end{itemize}

\item \textbf{Predictive incompleteness (option 3c)}
\vspace{-2mm}
\begin{itemize}
\item \textbf{Core idea:}   Quantum state is operationally incomplete until a measurement context is specified; the probabilities are Kolmogorovian \emph{per context}, not globally across incompatible observables. 
\item \textbf{Costs / risks:}  Requires explicit conditioning on context and non-Boolean event structure; forbids a single joint distribution for incompatible settings; challenges attitudes based on classical realism. 
\item \textbf{Benefits / implications:}  Preserves relativistic locality without superdeterminism; reframes Bell's inequalities violation as category errors of counterfactually mixing contexts; matches standard quantum formalism and experimental practice. 
\end{itemize}
\end{itemize}

\noindent To conclude this Appendix, we briefly sketch the arguments from \cite{entropy} to show the compatibility between predictive incompleteness and no‑signalling. 
Let's preserve  \emph{elementary locality}, also called parameter independence, that is the relativistic no‑signalling constraint,
and reject any superluminal causal influence from a distant setting to a local outcome. Bell's inequalities can still be violated, by rejecting \emph{predictive completeness}, also called outcome independence: the assumption that the operational state $\psi$ (possibly augmented by $\lambda$) alone suffices to determine the conditional probabilities of outcomes across incompatible contexts \cite{entropy}.  

Concretely, this asserts that the conditional probabilities relevant for Bell factorization must be written with explicit reference to the measurement context $C$,  
in the sense that the specification of $C$ is required to fix the distribution; therefore the factorization hypothesis used in Bell's theorem fails for the state $\psi$ alone. This failure is not a dynamical signalling mechanism: it is a statement about the insufficiency of $\psi$ to provide context‑independent outcome statistics, while parameter independence and no‑signalling remain intact.
\pagebreak

\end{document}